\documentstyle[amsfonts,amssymb]{article}

\newcommand{\mb}{\mathbb }
\newcommand{\beq}{\begin{eqnarray}}
\newcommand{\eeq}{\end{eqnarray}}
\newcommand{\beqst}{\begin{eqnarray*}}
\newcommand{\eeqst}{\end{eqnarray*}}

\newtheorem{theorem}{Theorem}[section]
\newtheorem{lemma}[theorem]{Lemma}
\newtheorem{corollary}[theorem]{Corollary}
\newtheorem{proposition}{Proposition}[section]

\title{\bf A Note on Wave Equation   in \\
Einstein \& de Sitter  Spacetime 
}

\author{{\bf Anahit Galstian,${}^{\small \mbox{\rm \scriptsize 1}}$ Tamotu Kinoshita,${}^{\scriptsize \mbox{\rm \scriptsize 2}}$ and Karen Yagdjian${}^{\scriptsize \mbox{\rm \scriptsize 3}}$ } }

\begin{document}

\date{}

\maketitle

\thispagestyle{empty}

\vspace*{-0.6cm}
\noindent
{\scriptsize{${}^1$\textsl{Department of Mathematics, 
University of Texas-Pan American,  1201 W.~University Drive,  Edinburg, TX 78541-2999, USA\\ 
{Electronic mail: agalstyan@utpa.edu }}\\
  ${}^2$\textsl{Institute of Mathematics, University of Tsukuba, 
 Tsukuba Ibaraki 305-8571, Japan \\ 
{Electronic mail: kinosita@math.tsukuba.ac.jp}}\\
 ${}^3$\textsl{Department of Mathematics, 
University of Texas-Pan American,  1201 W.~University Drive,  Edinburg, TX 78541-2999, USA\\
{Electronic mail: yagdjian@utpa.edu}}}}

\begin{abstract}
We consider the  wave propagating in the Einstein~\&~de~Sitter spacetime. 
The covariant d'Alembert's operator in the Einstein~\&~de~Sitter spacetime belongs
to the family of the non-Fuchsian partial differential operators.  
We introduce the   initial value problem for this equation and give the explicit representation
formulas for the solutions. We also show the $L^p - L^q$ estimates for solutions.

\medskip

\noindent
{\bf Keywords\,\, } Wave equation $ \cdot $   Einstein~\&~de~Sitter model   $ \cdot $    non-Fuchsian equations
\end{abstract}

\section{Introduction}

\setcounter{equation}{0}
\renewcommand{\theequation}{\thesection.\arabic{equation}}

The current note is concerned with the  wave propagating in the universe modeled by the cosmological
models with  expansion. We are motivated by the significant importance of
 the  solutions of the partial differential equations arising in the cosmological problems for our understanding of 
the universe.
While there exists  extensive literature on the hyperbolic equations, the question  of initial value problems
for the wave equation in the  curved spaces with singularities, and, in particular, in the Einstein~\&~de~Sitter  spacetime,  which are well posed and preserve
 many features of the classical waves, remains unresolved.
\medskip

The homogeneous and isotropic cosmological models possess highest symmetry, which  makes them  more amenable to rigorous study.
Among them,  FLRW (Friedmann-Lema$\hat {\mbox{i}}$tre-Robertson-Walker) models are mentioned. The simplest class of cosmological models 
can be obtained if we assume that the metric of the slices of constant time is flat and that 
the spacetime metric can be written in the form
\begin{eqnarray*}
ds^2= -dt^2+ a^2(t)( d{x}^2   + d{y} ^2 +  d{z}^2 )\,,
\end{eqnarray*}
with an appropriate scale factor $a(t)$. (See,e.g.,\cite{Peebles}.)
The assumption that the universe is expanding leads to the positivity of  
  the time derivative $\frac{d }{dt}a (t)$.  
The time dependence of the function $a(t)$ is  determined by  the Einstein field equations for gravity  
\[
R_{\mu \nu }-\frac{1}{2} g_{\mu \nu }R = -8\pi GT_{\mu \nu }. 
\] 
The metric of the Einstein~\&~de~Sitter universe (EdeS universe) is a particular member of the
Friedmann-Robertson-Walker metrics
\begin{eqnarray*}
ds^2= -dt^2+ a^2(t)\left[ \frac{dr^2}{1-Kr^2}   + r^2  d\Omega ^2 \right]\,,
\end{eqnarray*}
where $K=-1,0$, or $+1$, for a hyperbolic, flat or spherical spatial geometry, respectively. The Einstein equations
are simply
\begin{eqnarray}
&  &
\dot \rho =-3(\rho +p)\frac{\dot a}{a}\,, \qquad
\frac{\ddot  a}{a} =-\frac{4\pi }{3} (\rho +3p)\,, \qquad 
\label{Feq}
\left( \frac{\dot  a}{a} \right)^2 = \frac{8\pi }{3}  \rho  - \frac{K}{a^2}\,,
\end{eqnarray}
where $\rho $ is the proper energy density and $p $ is pressure.
For pressureless matter distributions ($p = 0$) and vanishing
spatial curvature ($K = 0$) in the EdeS universe the
solution to the  
equations (\ref{Feq}) is
\begin{eqnarray*}
&  &
a(t)= a_0 t^{{2}/{3}}\,,
\end{eqnarray*}
where $ a_0$ is an integration constant \cite{Ohanian-Ruffini}. This model describes
an open geometry (the $K = 0$ and spatial sections
are diffeomorphic to ${\mathbb R}^3$) in the presence of a constant
non-zero energy density distribution. Even though the
EdeS spacetime is conformally flat, its causal structure
is quite different from asymptotically flat geometries. In
particular, and unlike Minkowski or Schwarzschild, 
the past particle
horizons exist.
The EdeS spacetime is a good approximation to the
large scale structure of the universe during a matter
dominated phase, when the averaged (over space and time) energy density evolves adiabatically and pressures
are vanishingly small, as, e.g., immediately after inflation
[3–5]. This justifies why   such a metric  is adopted to
model the collapse of overdensity perturbations in the
early matter dominated phase that followed inflation.
\medskip

The   Einstein~\&~de~Sitter model  of the universe is the simplest non-empty expanding model with
the line-element
\[
ds^2 = - dt^2 + a_0^2t^{4/3} \left( dx ^2+ dy ^2+dz^2 \right)
\]
in comoving coordinates \cite{Ellis}.  It was first proposed jointly by Einstein~\&~de~Sitter (the EdeS model) \cite{Einstein-Sitter}.
The observations of the microwave radiation fit in with this model  \cite{Dirac}. 
The result of this case also correctly describes the early epoch, even in a universe with curvature different from zero \cite[Sec.~8.2]{Cheng}.
Recently it was used in \cite{Sultana-Dyer} to study cosmological black holes. The key observation for that approach is that the line-element 
can also be written  in the conformally flat form 
\[
ds^2 = \tau ^4[-d\tau ^ 2 + dr^ 2 + r^2(d\theta ^2 + \sin^2\theta  d \phi ^2)],  
\]
where the timelike coordinates are related by $ {d \tau }/{dt}
=  (3t)^{-{2}/{3}} $. The last form is  an asymptotic for the Schwarzschild metric whose line element may be written in the form 
\[
ds^2 = \tau ^4\left[ - \left( 1-\frac{2m}{r}\right)dt^2  + \frac{4m}{r}dt dr + \left( 1+\frac{2m}{r}\right)dr^2
+ r^2(d\theta ^2 + \sin^2\theta  d \phi ^2)\right],
\]
where $t$ and $r$ are timelike and spacelike coordinates related to the standard Schwarzschild coordinates $\bar t$ and $\bar r$, by 
\[
t=\bar t+2m \ln \left| \frac{\bar r}{2m}-1\right|, \quad r= \bar r.
\]
The fact that 
the resulting metric is asymptotically
Einstein~\&~de~Sitter with the source reducing to a comoving pure dust at
null infinity is  used in \cite{Sultana-Dyer}. In this sense the solution could be interpreted as a black hole in the asymptotic
background of the Einstein~\&~de~Sitter Universe.
\medskip

The covariant d'Alambert's operator in the Einstein~\&~de~Sitter spacetime is
\begin{eqnarray*}
\square_g \psi 
& = &
-  \left(  \frac{\partial}{\partial t} \right)^2  \psi 
+  t^{-{4}/{3}}  \sum_{i=1,2,3}\left( \frac{\partial }{\partial x^i} \right)^2   \psi  - \frac{2 }{ t }     \frac{\partial}{\partial t}    \psi \,.
\end{eqnarray*}
Consequently, the  covariant wave equation with the source term $f$ written in the coordinates  is
\begin{eqnarray}
\label{WE}
\left(  \frac{\partial}{\partial t} \right)^2  \psi 
-   t^{-{4}/{3}}  \sum_{i=1,2,3}\left( \frac{\partial }{\partial x^i} \right)^2   \psi +   \frac{2 }{ t }     \frac{\partial}{\partial t}    \psi =f\,.
\end{eqnarray}
The last equation belongs to the family of 
the non-Fuchsian partial differential equations.  There is very advanced theory of such equations (see, e.g.,
\cite{Mandai} and references therein). 
In this note we investigate the initial value problem for this equation and give the representation formulas for the solutions with any dimension $n \in {\mathbb N}$
of the spatial variable $x \in {\mathbb R}^n$.
\smallskip

The  equation (\ref{WE})  is   strictly hyperbolic in the domain with $t>0$. On the hypersurface  $t=0$ its coefficients  have   singularities  
that make   the study  of the initial value problem difficult. Then, the speed of propagation is  equal to $t^{-\frac{2}{3}} $    for every $ t \in {\mb R}\setminus \{0\}$. 
The  equation   (\ref{WE}) 
is not Lorentz  invariant, which brings additional difficulties. 
\smallskip

The classical works on the Tricomi  and  Gellerstedt  equations 
(see, e.g, \cite{Carroll-Showalter},  \cite{Delache-Leray}, \cite{Diaz-Weinberger}, \cite{Weinstein}) appeal
to the singular Cauchy problem for the Euler-Poisson-Darboux equation, 
\begin{eqnarray}
\label{E-P-D}
\Delta u = u_{tt} + \frac{c}{t}u_t \,, \qquad  c \in {\mb C}\,, 
\end{eqnarray}
and to the Asgeirsson mean value theorem
when handling a high-dimensional case. Here $\bigtriangleup  $ is the Laplace operator on the flat metric, 
$\bigtriangleup := \sum_{i=1}^n  \frac{\partial^2 }{\partial x_i^2}  $.

We use the approach suggested in \cite{YagTricomi} and reduce the problem for equation (\ref{WE}) to the Cauchy problem  for the 
free wave equation in Minkowski spacetime: 
$  v _{tt} -    \bigtriangleup  v     = 0$.
To  us, this approach seems to be more immediate than the one  that uses the Euler-Poisson-Darboux equation. More precisely, in the present note we   
utilize the solution $v=v(x,t;b)$ to the Cauchy problem
\begin{eqnarray}
\label{1.7new}\label{1.8new}
\cases{
v_{tt}-  \bigtriangleup v =0, \qquad t>0 ,\,\, x \in {\mathbb R}^n, \cr 
v(x,0)= \varphi  (x,b), \quad v_t(x,0)=0 ,\qquad x \in {\mathbb R}^n\,,}
\end{eqnarray}
with the parameter $b \in B \subseteq {\mb R}$. We  denote that solution by $v_\varphi =v_\varphi (x,t;b)$. In the case of 
function $\varphi $ independent of parameter,  we  skip $b$  and simply write $v_\varphi =v_\varphi (x,t)$.
There are well-known explicit representation formulas for the solution of the last problem. We write those formulas  to make the present note self-contained. 
If $n=1$, and $\varphi  (x,t)= f (x,t) \in C^\infty ({\mathbb   R}\times {\mathbb   R})$, $ B = {\mb R}$, then
\begin{eqnarray}
\label{1.10}
v_f (x,t;b) = \frac{1}{2} \left\{ f (x+t,b)+ f (x-t,b)\right\}, \qquad t \in {\mathbb R} ,\,\, x \in {\mathbb R} \,.
\end{eqnarray} 
For $f \in C^\infty ({\mathbb   R}^n\times {\mathbb   R})$ and for  odd $n=2m+1$, $m \in {\mathbb   N}$,
\begin{eqnarray}
\label{1.11}
 v_{f }  (x, t;b) =
 \frac{\partial}{\partial t} \Big( \frac{1}{t} \frac{\partial }{\partial t}\Big)^{\frac{n-3}{2} }
\frac{t^{n-2}}{\omega_{n-1} c_0^{(n)} } \int_{S^{n-1}  }
f  (x+ty,b)\, dS_y ,  \quad t \in {\mathbb R} ,\,\, x \in {\mathbb R}^n ,
\end{eqnarray}
where, $c_0^{(n)} =1\cdot 3\cdot 5\cdot \ldots\cdot (n-2)$, while for $x \in {\mathbb   R}^n$ with even $n=2m$,  $m \in {\mathbb   N}$,
\begin{eqnarray}
\label{1.12}
v_{f }  (x, t;b) =   \frac{\partial }{\partial t}
\Big( \frac{1}{t} \frac{\partial }{\partial t}\Big)^{\frac{n-2}{2} }
\frac{2t^{n-1}}{\omega_{n-1} c_0^{(n)}} \int_{B_1^{n}(0)}  \frac{1}{\sqrt{1-|y|^2}}f (x+ty,b)\, dV_y ,\quad t \in {\mathbb R} ,\,\, x \in {\mathbb R}^n,
\end{eqnarray}
where, $c_0^{(n)} = 1\cdot 3\cdot 5\cdot \ldots\cdot (n-1)$. (See, e.g. Theorems~4.1,4.2~\cite{Shatah}.) In particular, if $f$ is independent of $t$, then
$v_f(x,t;b)$ does not depend on $b$ and we briefly write $v_f(x,t)$.
\medskip

The straightforward application of the formulas obtained in   \cite{YagTricomi}
to the Cauchy problem for equation (\ref{WE}) decidedly does not work, but it reveals a surprising link to the Einstein~\&~de~Sitter spacetime.  
To demostrate that link we note that the ``principal part'' of equation (\ref{WE})
belongs to the family of the Tricomi-type equations (in the case of odd $l$ it is Gellerstedt equation):
\[ 
u_{tt} - t^{l} \bigtriangleup u =  0 \,, 
\]
where $l \in {\mathbb N}$. According to \cite{YagTricomi} the solution to the Cauchy problem  
\[
u_{tt}-  t^{l} \bigtriangleup u =f(x,t), \quad u(x,0)= \varphi_0 (x), \quad u_t(x,0)=\varphi_1 (x),
\]
with the smooth functions $ f$, $ \varphi_0 $,  and  $\varphi_1  $, can be represented  as follows:
\begin{eqnarray}
\label{solution}
u(x,t) 
& = &
 2^{2-2\gamma }    
\frac{\Gamma \left( 2\gamma  \right) } {\Gamma^2 \left( \gamma  \right) }  
\int_{0}^1  (1-s^2)^{\gamma - 1   } v_{\varphi _0 } (x, \phi (t) s) 
 ds \\
&  &
+\, t 2^{2\gamma }   
\frac{\Gamma \left(2- 2\gamma  \right) } {\Gamma^2 \left( 1- \gamma  \right) }
\int_{0}^1  (1-s^2)^{- \gamma  }  v_{\varphi _1 }  (x, \phi (t) s) 
ds  \nonumber\\
&   &
+\, 2c_k \int_{ 0}^{t} db \!\! \int_{ 0 }^{ \phi (t) - \phi (b)} \!\!
dr\,  E(r,t;0,b)  v_{f } (x,r;b)\, , \quad x \in {\mathbb R}^n, \,\, t>0\,,   \nonumber
\end{eqnarray}
with the kernel 
\beq
\label{Gel2.8a}
\hspace*{-0.5cm} E(r,t;0,b)
& := &
 (  (\phi (b)  + \phi (t))^2 -r^2  )^{-\gamma} F \left(\gamma, \gamma;1; \frac{ ( \phi (t)  - \phi (b))^2 -r^2 }
{( \phi (t)  + \phi (b))^2 -r^2 }  \right)\,.
\eeq
Here $F\big(\gamma, \gamma;1; \zeta \big) $ is the hypergeometric function (see, e.g., \cite{B-E}),  
while\,  $k :=l/2$, $ \phi (t):= \frac{t^{k+1} }{k+1}$, $ \gamma := \frac{k}{2k+2} $, and  $c_k = (k+1)^{-\frac{k}{k+1} }2^{- \frac{1}{k+1} }$.
The equation with $l \in {\mathbb N}$, $x \in {\mathbb R}$, and $f=0$ is studied in \cite{T-T} by means of the partial Fourier transform and the confluent 
hypergeometric function. That approach gives parametrix of the Cauchy problem  and as a consequence, a  complete description of the propagation of the $C^\infty$-singularities.
\medskip

Suppose now that we are looking for the simplest possible kernel $E(r,t;0,b)$ (\ref{Gel2.8a}) of the last integral transform. In the hierarchy of the  hypergeometric functions
the simplest one, that is different from the constant, is a linear function. That simplest function $F \left(a, b;1; \zeta  \right) $ has the parameters $a=b=-1$
and coincides with $1+ \zeta $. The parameter $l$ leading to  such function  $F \left(-1, -1;1; \zeta  \right) = 1+ \zeta $ is exactly the exponent $l=-4/3$ of the 
wave equation (and of the metric tensor) in the Einstein~\&~de~Sitter spacetime. 
\medskip

It is evident that the first term of the representation (\ref{solution}), as it is written, is meaningless if $\gamma =-1$. This indicates the fact that the Cauchy problem is not 
well-posed anymore for the equation with $l=-4/3$. The next theorem also shows how the ``lower order term'' of the equation (\ref{WE}) affects the Cauchy  problem.
The main result of this paper is the following theorem.
\begin{theorem}
\label{T1.1}
Assume that $\varphi_i  \in C^{[\frac{n}{2}]+3-i} ({\mathbb  R}^n) $, $i=0,1$, 
$ f(x,t) \in C^{[\frac{n}{2}]+2} ({\mathbb  R}^n\times (0,\infty))$, and that with some $\varepsilon >0$ one has
\[
|\partial_x^\alpha f(x,t)| + |t \partial_t \partial_x^\beta  f (x,t)| \leq C_\alpha t^{\varepsilon -2 }  \quad \mbox{for all}  \,\, x \in {\mathbb R}^n, 
\quad    \mbox{and for all small}\,\, t>0,
\]
and for every $\alpha  $, $\beta  $, $|\alpha | \leq  [\frac{n}{2}]+2$, $|\beta  | \leq  [\frac{n}{2}]+1$.  Then the solution $ \psi = \psi (x,t)$ to the problem  
\begin{eqnarray}
\label{ivp}
\begin{cases}{
 \psi_{tt} - t^{-4/3}\bigtriangleup    \psi +   2   t^{-1}          \psi_t = f (x,t),  \qquad t>0 ,\,\, x \in {\mathbb R}^n,\cr 
 \displaystyle   \lim_{t\rightarrow 0}\, t \psi  (x,t) = \varphi_0 (x), \quad   
\displaystyle   
\lim_{t\rightarrow 0} 
  \left(  t \psi_t  (x,t) + \psi  (x,t)+3 t^{ - {1}/{3}} \bigtriangleup \varphi_0   (x   )  \right)
=  \varphi_1 (x),   \,\, x \in {\mathbb  R}^n , }
\end{cases}
\end{eqnarray} 
is given by 
\begin{eqnarray}
\label{1.15}
\psi (x,t)
& = & 
\frac{3}{2} t^2 \int_{ 0}^{1} db
\int_{ 0 }^{ 1 - b^{1/3} }  ds \,  bv_f (x,3t^{1/3}  s ; tb)  
 \big(1   + b^{2/3}-s^2  \big) \\
&  &
+ \, t^{  -1  } v_{\varphi_0}  (x, 3t^{1/3})
- 3t^{-2/3}  \left(  \partial_t   v_{\varphi_0}  \right)  (x, 3t^{1/3})  +
 \frac{3}{2} 
\int_{0}^1   v_{\varphi _1 } (x, 3t^{1/3} s) 
(1-s^2) ds \,.\nonumber 
\end{eqnarray}
\end{theorem}
The theorem shows that one cannot anticipate the well-posedness in the Cauchy problem for the 
wave equation in the Einstein~\&~de~Sitter spacetime.  In fact, it gives a structure of singularity  of the solution at the point $t=0$, which  
 hints at the proper initial conditions which have to be
prescribed for the solution.
The initial conditions prescribed in  the previous theorem are the Cauchy conditions modified 
 to the so-called {\it weighted initial conditions}  in order 
to adjust them to the equation. For the Euler-Poisson-Darboux equation (\ref{E-P-D}) one can find such weighted initial conditions,
for instance, in books \cite{Carroll-Showalter} and \cite{Smirnov} as well as in the references therein. The existence and uniqueness of the solutions for 
the initial value problem with 
the weighted initial conditions  for the Euler-Poisson-Darboux equation
and for the equation (\ref{E-P-D}) with the time-dependent $c$ are 
proved in \cite{DelSanto-Kinoshita-Reissig} by application of the Fourier transform  in $x$-variable,
as well as some transformations which 
reduce the equation to the confluent hypergeometric equation. 
\medskip

Theorem~\ref{T1.1} can be used to obtain  some important properties for the solutions of the wave 
equation in Einstein~\&~de~Sitter spacetime, 
which are inherited from the  solutions of the wave equation in Minkowski spacetime. 
In particular, as a consequence  of the previous theorem,  in Section~\ref{S3}, 
for the initial value problem (\ref{ivp}) with $n\geq 2$, $f=0$, and $\varphi _0 =0$,  we obtain
the following $L^p-L^q $ estimate 
\begin{eqnarray*} 
\| (-\bigtriangleup )^{-s} \psi (\cdot ,t)\|_{ { L}^{q} ({\mathbb R}^n)  }
& \leq  &  
C t^{\frac{1}{3}\left( 2s-n(\frac{1}{p}-\frac{1}{q}) \right)}\|\varphi_1   \|_{ { L}^{p}({\mathbb R}^n)  }, \quad \, t>0\,,
\end{eqnarray*}
provided that $s\ge 0$, $1<p\le 2$, $\frac{1}{p}+\frac{1}{q}=1$, $\frac{1}{2}(n+1)(\frac{1}{p}-\frac{1}{q})   \le 2s 
\le n (\frac{1}{p}-\frac{1}{q})$,  and
$n(\frac{1}{p}-\frac{1}{q})  -1 < 2s $. Similar estimates hold for the problem with general $\varphi _0 $ and $f $. 
Thus, in the present paper we prepare all necessary tools that will allow us 
to study in the forthcoming paper the solvability of semilinear wave equation 
in the Einstein~\&~de~Sitter spacetime. Having in mind the scale invariance of the equation and also the 
applications (see, e.g.,  \cite{Henriksen-Wesson}, \cite{Suginohara-Taruya-Suto}),  special attention will be given to the self-similar solutions. 
Results analogous to those  presented in this note have already proven to be a good tool in the study of
self-similar solutions \cite{YagTricomi_JMAA}.
\medskip

This note is organized as follows. In Section \ref{S2} we prove the main theorem and give some of its extensions (Theorems~\ref{T2.1new}-\ref{T2.2new}) that allow 
stronger singularity in the source term.  Section \ref{S3} is devoted to the
 application of the main theorem, namely, to the derivation of the $L^p-L^q$ estimates.  
\medskip

The EdeS model recently became a focus of interest for an increasing 
number of authors. (See, e.g., \cite{Blanchard}, \cite{Ellis},  \cite{Goncalves}, \cite{Gron-Hervik}, \cite{Hawking}, \cite{Hawley},
\cite{Peebles}, \cite{Rendall_book}, \cite{Sultana-Dyer} and references therein.) We believe that the initial value problem and 
the explicit representation formulas obtained in the present paper fill the gap in the existing literature on the wave equation in the EdeS 
spacetime.

\section{Proof of the main theorem}
\setcounter{equation}{0}
\renewcommand{\theequation}{\thesection.\arabic{equation}}
\label{S2}

If we denote
\[
{\mathcal L} :=\partial_{t}^2
-  t^{-\frac{4}{3}} \bigtriangleup      + 2 t^{-1 }        \partial_t  ,\qquad 
{\mathcal S}:= \partial_{t}^2
-  t^{-\frac{4}{3}} \bigtriangleup   \,,    
\]
then we can easily check for $t \not= 0$ the following operator identity 
\begin{equation}
 t^{-1} \circ {\mathcal S}  \circ t   
 =  
   {\mathcal L}   \,.
\end{equation}
The last equation suggests  a change of unknown function $\psi $ with $u$ such that $\psi =t^{-1}u $. Then the problem for $u$ is as follows: 
\begin{eqnarray}
\label{2.2a}
\begin{cases}{
\vspace{0.2cm}   u_{tt} - t^{-4/3} \bigtriangleup    u  = g(x,t),  \qquad t>0 ,\,\, x \in {\mathbb R}^n,\cr 
 \displaystyle   \lim_{t\rightarrow 0}\, u (x,t) = \varphi_0 (x), \qquad  \,\, x \in {\mathbb R}^n, \cr 
\displaystyle   
\lim_{t\rightarrow 0} 
  \left(  u _t (x,t)+3 t^{ - {1}/{3}} \bigtriangleup  \varphi_0  (x   )  \right)
=  \varphi_1 (x),   \quad x \in {\mathbb  R}^n ,} 
\end{cases}
\end{eqnarray}
where $g(x,t)=tf(x,t)$. 
Therefore it is enough to find a representation of the solution of the last problem. We  discuss it in three
separate cases of: {\bf ($\bf f$)} with  $\varphi_0 =\varphi _1=0$; {\bf ($\bf \varphi_0$)} with  $f=0$ and $\varphi _1=0$; {\bf ($\bf \varphi_1$)} with  $f=0$ and $\varphi _0=0$.

We will use the following property of the resolving operator of the problem (\ref{1.7new}): if $P=P(D_x) $ is a pseudo-differential operator  and $\tau =\tau (t)$ is a  
smooth function
of time, then the operator $\tau (t) P(D_x) $ ``commutes'' with the resolving operator of the problem (\ref{1.7new}). More precisely, the following identity can easily be verified:
\begin{equation}
 v_{\tau P(D_x)f}(x,t;b) = \tau (b) P(D_x) v_{ f}(x,t;b) \quad \mbox{\rm for all}\quad f  \in C^\infty ({\mathbb  R}^n\times (0,\infty))\,.
\end{equation}

The operator ${\mathcal S} $ belongs to the family of the Tricomi-type operators 
\[
{\mathcal T}:= \partial_{t}^2
-  t^{l} \bigtriangleup\,,
\]
where $l \in {\mathbb R}$. The Cauchy problem for such operators with  positive $l$, that is for the case of multiple characteristics,
 is well developed (see, e.g., \cite{YagBook} and references therein). 
The fundamental solutions of the operator and the representation formulas for the solutions of the Cauchy problem are given
in \cite{YagTricomi}. The results of \cite{YagTricomi} allow us to write an {\it ansatz} for the solutions of the equation of (\ref{2.2a}). This {\it ansatz}  has been used in 
\cite{Kinoshita-Yagdjian} to extend the range of admissible   values of $l$ to negative numbers  for the problem with data on the hyperplane $t\not= 0$.
Here we use this {\it ansatz}  to consider the weighted initial value problem  (\ref{2.2a}) with data on the plane $t=0$, where coefficients and source term 
are singular. As we already emphasized, it is interesting that the case of $l=-4/3$, that is the case of Einstein~\&~de~Sitter spacetime, is an exceptional case in
the sense that it simplifies the Gauss' hypergeometric function $F(\gamma ,\gamma ;1;z)$ appearing in the fundamental solutions constructed in \cite{YagTricomi}, to
the linear function $F(-1 ,-1 ;1;z)=1+z$.
\bigskip

\noindent
{\bf The case of {\bf ($\bf f$)}.} 
Assume that $f \in C^{[\frac{n}{2}]+2}({\mathbb R}^n\times (0,\infty)  )$ and for every given multiindexes $\alpha $, $\beta $, $|\alpha | \leq [\frac{n}{2}] +2$, 
$|\beta  | \leq [\frac{n}{2}] +1$,
the following inequality holds
\[
|\partial_x^\alpha  f(x,t)| +|t\partial_t \partial_x^\beta  f(x,t)| \leq Ct^{\varepsilon-2}  \quad \mbox{ for all} \,\, x \in {\mathbb R}^n ,
 \quad    t\in (0,T) ,
\] 
for small positive $T$. We have to prove that   the solution to the problem 
\[
\cases{ 
   {\mathcal L} \psi   = f,  \quad t>0,\,\, x \in {\mathbb R}^n, \cr
 \displaystyle   \lim_{t\rightarrow 0}\psi   (x,t) =0,  \quad 
 \displaystyle   \lim_{t\rightarrow 0}
\psi _t  (x,t) 
=0 , \quad   x \in {\mathbb R}^n,  } 
\]
is given by 
\begin{eqnarray}
\psi (x,t)
& = & 
\frac{3}{2} t^2 \int_{ 0}^{1} db
\int_{ 0 }^{ 1 - b^{1/3} }  ds \,  bv_f (x,3t^{1/3}  s ; tb)  
 \big(1   + b^{2/3}-s^2  \big) \,.
\end{eqnarray}
Here the function $v_f (x,r ; t ) $ is given by (\ref{1.10}), (\ref{1.11}), and (\ref{1.12}), if $n=1$, $n$ is odd, and $n$ is even, respectively.

\medskip

It is sufficient to check the properties of the function $u=u(x,t)=t\psi (x,t)$, which solves the equation
$ {\mathcal S} u   = g$ with $g(x,t)=tf(x,t)$. Hence, we can restrict ourselves to the representation
\begin{eqnarray}
\label{2.5}
u(x,t)
& = & 
\frac{3}{2} t^2 \int_{ 0}^{1} db
\int_{ 0 }^{ 1 - b^{1/3} }  ds \,  v_{g} (x, \phi (t)  s ; tb)  
 \big(1   + b^{2/3}-s^2  \big) \nonumber \\
& = & 
 \frac{1}{18}   \int_{ 0}^{t} dl
\int_{ 0 }^{ \phi (t) - \phi (l) }  dr \,  v_{g} (x,r ; l)  
 \left(\phi^2 (t)  + \phi^2 (l) -r^2  \right) 
\end{eqnarray}
and take into account the identity $v_{g} (x,r ; b) = b v_{f} (x,r ; b)$. Here $\phi (t):= 3t^{1/3}$.
First we prove that the integral is convergent and that it represents a $C^2 ({\mathbb R}^n\times (0,\infty) )$-function. 
 We will skip the subindex $g$ in the remainder of the proof.
It is evident that $v (x,r ; b) \in C^2 ({\mathbb R}^n\times (0,\infty) )$ and that 
\begin{eqnarray*}
|\partial_x^\alpha  v (x,r ; b)|  +|  \partial_r v(x,r ; b)|  +|  \partial_r^2 v(x,r ; b)| + | b \partial_b  v  (x,r ; b)| \leq Cb^{-1+\varepsilon}\,,
\end{eqnarray*}
if  $|\alpha | \leq 2 $. It follows 
\begin{eqnarray*}
| v (x,\phi (t)  s; tb)| \leq    Ct^{-1+\varepsilon}b^{-1+\varepsilon} \,.
\end{eqnarray*}
Then we use the last inequality and the first formula of (\ref{2.5}) in the following inequalities: 
\begin{eqnarray}
| u(x,t) |
& \leq  & 
 Ct^{1+\varepsilon}   \int_{ 0}^{1} b^{-1+\varepsilon} db
\int_{ 0 }^{ 1 - b^{1/3} }  ds \,   
 \big(1   + b^{2/3}-s^2  \big)  \nonumber \\
 \label{uiv}
 & \leq &
 C_\varepsilon t^{1+\varepsilon}  \,.
\end{eqnarray}
The first formula of (\ref{2.5}) leads to the estimate for the derivative
\begin{eqnarray*}
\left| \frac{\partial }{\partial t} u(x,t) \right|
 & \leq &
\left|  3 t  \int_{ 0}^{1} db
\int_{ 0 }^{ 1 - b^{1/3} }  ds \,   v (x,\phi (t)  s ; tb)  
 \big(1   + b^{2/3}-s^2  \big)  \right| \\
&  &
+ \left| \frac{3}{2}   t^{4/3}  \int_{ 0}^{1} db
\int_{ 0 }^{ 1 - b^{1/3} }  ds \,  s (\partial_r v)  (x,\phi (t)  s ; tb)  
 \big(1   + b^{2/3}-s^2  \big)  \right|\\
&  &
+ \left| \frac{3}{2} t^2  \int_{ 0}^{1} db
\int_{ 0 }^{ 1 - b^{1/3} }  ds \,   b (\partial_b v)   (x,\phi (t)  s ; tb)  
 \big(1   + b^{2/3}-s^2  \big)  \right| 
\end{eqnarray*}
that implies
\begin{eqnarray}
\left| \frac{\partial }{\partial t} u(x,t) \right|
 & \leq &
C_\varepsilon    t^{\varepsilon }\int_{ 0}^{1} b^{-1+\varepsilon }db
\int_{ 0 }^{ 1 - b^{1/3} }  ds \,  
 \big(1   + b^{2/3}-s^2  \big)   \nonumber \\
&  &
+ C_\varepsilon  t^{1/3+\varepsilon}   \int_{ 0}^{1} b^{-1+\varepsilon } db
\int_{ 0 }^{ 1 - b^{1/3} }  ds \,  s  
 \big(1   + b^{2/3}-s^2  \big)  \nonumber \\
 \label{uivder}
&  &
+ C_\varepsilon  t^\varepsilon  \int_{ 0}^{1}  b^{-1+\varepsilon} db
\int_{ 0 }^{ 1 - b^{1/3} }  ds \,   
 \big(1   + b^{2/3}-s^2  \big)  \,.
\end{eqnarray}
Thus, the estimates (\ref{uiv}) and (\ref{uivder}) lead to the initial conditions
\begin{equation}
 \displaystyle  \lim_{t\rightarrow 0} u (x,t) =0,  \quad 
 \displaystyle   \lim_{t\rightarrow 0}
u_t  (x,t) 
=0 .  
\end{equation}
It remains to verify the equation. For the derivative $\frac{\partial }{\partial t} u(x,t) $ we use the second formula of (\ref{2.5}) and
obtain
\begin{eqnarray*}
\frac{\partial }{\partial t} u(x,t)
& = & 
 \phi' (t)    \frac{1}{18}   \int_{ 0}^{t} v (x,\phi (t) - \phi (l) ; l)  
 \Big(\phi^2 (t)  + \phi^2 (l) -(\phi (t)  -\phi (l)) ^2  \Big)   \, dl\\
 &  &
 + 2\phi' (t) \phi (t) \frac{1}{18}   \int_{ 0}^{t} dl
\int_{ 0 }^{ \phi (t) - \phi (l) }  dr \,  v (x,r ; l) \\
& = & 
\frac{1}{18}   \Big( \phi^2 (t) \Big)'  \int_{ 0}^{t} v (x,\phi (t) - \phi (l) ; l)  
 \phi  (l)   \, dl 
 +   \frac{1}{18}\Big( \phi^2 (t) \Big)'   \int_{ 0}^{t} dl
\int_{ 0 }^{ \phi (t) - \phi (l) }  dr \,  v (x,r ; l) \,.
\end{eqnarray*}
For the second order derivative  $\frac{\partial^2}{\partial t^2} u(x,t)$,
since $\frac{1}{18}(\phi^2(t))' v(x,0;t)\phi (t)=g(x,t)$ we derive from the last equation
\begin{eqnarray}
\label{u_tt}
&  &
\frac{\partial^2 }{\partial t^2} u(x,t) \nonumber \\
& = & 
g(x,t) \nonumber  \\
 &  & 
+      \frac{1}{18}    ( \phi^2 (t)  )'{ }'  \int_{ 0}^{t} v (x,\phi (t) - \phi (l) ; l)  
 \phi  (l)   \, dl 
+   \frac{1}{18}    ( \phi^2 (t)  )' \phi' (t) \int_{ 0}^{t} v _r(x,\phi (t) - \phi (l) ; l)  
 \phi  (l)   \, dl   \nonumber \\
&  & 
+  \frac{1}{18} ( \phi^2 (t)  )'{ }'    \int_{ 0}^{t} dl
\int_{ 0 }^{ \phi (t) - \phi (l) }  dr \,  v (x,r ; l)  +   \frac{1}{18} ( \phi^2 (t)  )'   \phi' (t) \int_{ 0}^{t} v (x,\phi (t) - \phi (l)  ; l)   \,dl \,. 
\end{eqnarray}
By means of the second formula of (\ref{2.5}) and equation of (\ref{1.7new}) for the   function  $\bigtriangleup u  $ we derive
\begin{eqnarray*}
\bigtriangleup  u(x,t)
& = & 
\frac{1}{18}   \int_{ 0}^{t} dl
\int_{ 0 }^{ \phi (t) - \phi (l) }  dr \,  \partial_r^2 v  (x,r ; l)  
 \left(\phi^2 (t)  + \phi^2 (l) -r^2  \right).
\end{eqnarray*}
It follows
\begin{eqnarray*}
\bigtriangleup  u(x,t)
& = & 
\frac{1}{9}   \phi  (t) \int_{ 0}^{t} 
   \partial_r v (x,\phi (t) - \phi (l)  ; l)   \phi (l) \,dl  \\
&  &
- \frac{1}{18}   \int_{ 0}^{t} 
  \partial_r v (x,0  ; l) 
 \left(\phi^2 (t)  + \phi^2 (l)   \right)\, dl
 + \frac{1}{9}   \int_{ 0}^{t} dl
\int_{ 0 }^{ \phi (t) - \phi (l) }  dr \,  r   \partial_r v  (x,r ; l)\,.
\end{eqnarray*}
Since $\partial_r v(x,0;l)=0 $, one more integration by parts yields
\begin{eqnarray}
\label{uxx}
\bigtriangleup  u(x,t)
& = & 
\frac{1}{9}   \phi  (t) \int_{ 0}^{t} 
  \partial_r v (x,\phi (t) - \phi (l)  ; l)   \phi (l) \,dl  \\
 &  &
 + \frac{1}{9}   \int_{ 0}^{t}  v  (x,\phi (t) - \phi (l) ; l) (\phi (t) - \phi (l))  \,  dl  
- \frac{1}{9}   \int_{ 0}^{t} dl \int_{ 0 }^{ \phi (t) - \phi (l) }   dr  \,v (x,r ; l)  \,. \nonumber
\end{eqnarray}
Hence according to (\ref{u_tt}) and (\ref{uxx}) the application of the operator\, $\partial_t^2- t^{-4/3} \bigtriangleup  $\, to the  function \,$u=u(x,t) $ \, gives
\begin{eqnarray*}
&  &
  u_{tt}(x,t)  - t^{-4/3} \bigtriangleup  u(x,t) \\
& = &
g(x,t)  \\
 &  &  
+      \frac{1}{18}    ( \phi^2 (t)  )'{ }'  \int_{ 0}^{t} v (x,\phi (t) - \phi (l) ; l)  
 \phi  (l)   \, dl 
+   \frac{1}{18}    ( \phi^2 (t)  )' \phi' (t) \int_{ 0}^{t} v _r(x,\phi (t) - \phi (l) ; l)  
 \phi  (l)   \, dl  \\
&  & 
 + \frac{1}{18} ( \phi^2 (t)  )'   \phi' (t) \int_{ 0}^{t} v (x,\phi (t) - \phi (l)  ; l)   \,dl \\
&  &
-   t^{-4/3} \Bigg\{ \frac{1}{9}   \phi  (t) \int_{ 0}^{t} 
   v_{r } (x,\phi (t) - \phi (l)  ; l)   \phi (l) \,dl  
 + \frac{1}{9}   \int_{ 0}^{t}  v  (x,\phi (t) - \phi (l) ; l) (\phi (t) - \phi (l))  \,  dl    \Bigg\}\\
& = &
g(x,t) \\
 &  &  
+      \frac{1}{18}    ( \phi^2 (t)  )'{ }'  \int_{ 0}^{t} v (x,\phi (t) - \phi (l) ; l)  
 \phi  (l)   \, dl  
 + \frac{1}{18} ( \phi^2 (t)  )'   \phi' (t) \int_{ 0}^{t} v (x,\phi (t) - \phi (l)  ; l)   \,dl \\
&  &
-   t^{-4/3}  \frac{1}{9}   \int_{ 0}^{t}  v  (x,\phi (t) - \phi (l) ; l) (\phi (t) - \phi (l))  \,  dl  \\
& = &
g(x,t) 
 + \frac{1}{18} ( \phi^2 (t)  )'   \phi' (t) \int_{ 0}^{t} v (x,\phi (t) - \phi (l)  ; l)   \,dl 
-   t^{-4/3} \frac{1}{9}   \phi (t)  \int_{ 0}^{t}  v  (x,\phi (t) - \phi (l) ; l) \,  dl  \\
& = &
g(x,t) \,.
\end{eqnarray*}
Thus, for this case the theorem is proven. 
\medskip

One can allow more strong singularity of the source function $f$ at $t=0$. 
More precisely, one can reduce the case with such singularity  
to the one of Theorem~\ref{T1.1}   if the initial condition are modified. That is done in the next theorem.

\begin{theorem}
\label{T2.1new}
Assume that 
$ f(x,t) \in C^{ [\frac{n}{2}]+4} ({\mathbb  R}^n\times (0,\infty))$, $ t^2 f(x,t) \in C  ({\mathbb  R}^n\times [0,\infty))$, and that  
\[
|\partial_x^\alpha f(x,t)| + |t \partial_t\partial_x^\beta  f(x,t)| \leq C_\alpha t^{ -2 }  \quad \mbox{for all}\quad  t\in (0,T) ,\,\, x \in {\mathbb R}^n, 
\]
and for every $\alpha  $, $\beta  $, $|\alpha | \leq  [\frac{n}{2}]+4$, $|\beta  | \leq  [\frac{n}{2}]+3$.  Denote $\displaystyle f_0(x):= \lim_{t \to 0} t^2 f(x,t)$ 
and suppose that   with some $\varepsilon >0$ for the functions $f(x,t) $   and $f_0 (x) \in C^{ [\frac{n}{2}]+4} ({\mathbb  R}^n)$
the following inequality is fulfilled:
\[
|\partial_x^\alpha \left( t  f(x,t) -t^{-1}f_0(x) \right) | + |t \partial_t \partial_x^\beta  \left( t  f(x,t) -t^{-1}f_0(x) \right)| \leq C_\alpha t^{\varepsilon  -1 }\,\, 
\mbox{for all}\,\,   t\in (0,T) ,\,\, x \in {\mathbb R}^n, 
\]
and for every $\alpha  $, $\beta  $, $|\alpha | \leq  [\frac{n}{2}]+2$, $|\beta  | \leq  [\frac{n}{2}]+1$.

Then the solution $\psi =\psi (x,t) $ of the problem  
\begin{eqnarray}
\label{2.4CP}
\begin{cases}{
\vspace{0.2cm}  \psi_{tt} - t^{-4/3}\bigtriangleup    \psi +   2   t^{-1}          \psi_t = f (x,t),  \qquad  t>0  ,\,\, x \in {\mathbb R}^n,\cr 
 \displaystyle   \lim_{t \to 0} \left( t \psi   (x,t) \right)
  =     
0 , \qquad 
 \lim_{t \to 0} \left(  t \psi _t (x,t)+ \psi   (x,t)  - f_0(x)\ln t \right)
  =  0 \,, \quad x \in {\mathbb  R}^n , }
\end{cases}
\end{eqnarray} 
is given by 
\begin{eqnarray}
\label{2.4}
\psi  (x,t) 
& = & 
\frac{1}{t} f_0(x)\tau (t )+ \frac{1}{18t}   \int_{ 0}^{t} dl
\int_{ 0 }^{ \phi (t) - \phi (l) }  dr \, 
 (\phi^2 (t)  + \phi^2 (l) -r^2  ) \\
&  &
\hspace{3.2cm}  \times \left( l v_{ f }(x,r;l) -l^{-1}v_{f_0 } (x,r)+ l^{-4/3}\tau (l)\bigtriangleup  v_{  f_0  }(x,r)\right),\nonumber 
\end{eqnarray}
where $\tau (t):= \int_0^t \ln s \,ds $.
\end{theorem}
\medskip

\noindent
{\bf Proof.} Consider the new unknown function $ w(x,t):=u-f_0(x)\tau (t ) $. 
Then
\begin{eqnarray*}
{\mathcal S}w(x,t)
& = &
tf(x,t) - {\mathcal S}(f_0(x)\tau (t) ) \\
& = &
tf(x,t)    -\left(t^{-1}f_0(x)-t^{-4/3}\tau (t)  \bigtriangleup    f_0(x) \right)\\
& = &
h(x,t),
\end{eqnarray*}
where we have denoted   
\[
h(x,t) := t\left( f(x,t) -t^{-2}f_0(x)\right) + t^{-4/3}\tau (t) \bigtriangleup    f_0(x)\,.
\]
According to the condition of the theorem with some $\varepsilon >0 $ we have
\[
|\partial_x^\alpha h(x,t)| + |t \partial_t \partial_x^\beta  h(x,t)| \leq C_\alpha t^{\varepsilon  -1 }  \qquad \mbox{for all}\quad   t\in (0,T) ,\,\, x \in {\mathbb R}^n, 
\]
$\alpha $, $\beta  $, $|\alpha | \leq [\frac{n}{2}]+2$, $|\beta  | \leq [\frac{n}{2}]+1$, that allows us to write representation (\ref{2.5}) for the solution $w= w(x,t)$:
\begin{eqnarray}
\label{2.13}
w(x,t)
& = & 
 \frac{1}{18}   \int_{ 0}^{t} dl
\int_{ 0 }^{ \phi (t) - \phi (l) }  dr \,  v_{h} (x,r ; l)  
 (\phi^2 (t)  + \phi^2 (l) -r^2  )\,.
\end{eqnarray}
On the other hand, according to Theorem~\ref{T1.1}, the function $w=w(x,t) $ satisfies initial conditions 
\begin{eqnarray}
\label{2.14}
\lim_{t \to 0} w  (x,t)
& = &   
0, \qquad \lim_{t \to 0}  w_t (x,t)
  =     0 \,.
\end{eqnarray}
Consequently,
\begin{eqnarray*}
\lim_{t \to 0} u  (x,t)
& = &   
\lim_{t \to 0} ( w  (x,t)  + f_0(x)\tau (t )) =0 , \\
 \lim_{t \to 0} ( u_t (x,t)- f_0(x)\ln t )
 & =  &
\lim_{t \to 0}  w_t  (x,t) =  0 \,.
\end{eqnarray*}
For the function $\psi =\psi (x,t) = t^{-1}u (x,t) $ this implies the initial conditions of (\ref{2.4CP}). 
To prove representation formula (\ref{2.4}), we note that
\begin{eqnarray*}
v_h (x,r;b) 
& = &
v_{t\left( f(x,t) -t^{-2}f_0(x)\right) + t^{-4/3}\tau (t)  \bigtriangleup   f_0(x)}(x,r;b)\\
& = &
v_{t   f(x,t) }(x,r;b) -v_{t^{-1}f_0(x)} (x,r;b)+ v_{t^{-4/3}\tau (t)  \bigtriangleup   f_0(x) }(x,r;b)\\
& = &
b v_{ f }(x,r;b) -b^{-1}v_{f_0 } (x,r;b)+ b^{-4/3}\tau (b)  \bigtriangleup   v_{f_0  }(x,r;b)\\
& = &
b v_{ f }(x,r;b) -b^{-1}v_{f_0 } (x,r)+ b^{-4/3}\tau (b)  \bigtriangleup   v_{f_0  }(x,r)\,.
\end{eqnarray*}
Then we use (\ref{2.13}) to write
\begin{eqnarray*}
w(x,t) 
& = &  
 \frac{1}{18}   \int_{ 0}^{t} dl
\int_{ 0 }^{ \phi (t) - \phi (l) }  dr \, 
 (\phi^2 (t)  + \phi^2 (l) -r^2  ) \\
&  &
\qquad \times \left( l v_{ f }(x,r;l) -l^{-1}v_{f_0 } (x,r)+ l^{-4/3}\tau (l) \bigtriangleup  v_{f_0  }(x,r)\right)    \,.
\end{eqnarray*}
Thus, the representation
\begin{eqnarray*}
u (x,t) 
& = & 
f_0(x)\tau (t )+ \frac{1}{18}   \int_{ 0}^{t} dl
\int_{ 0 }^{ \phi (t) - \phi (l) }  dr \, 
 (\phi^2 (t)  + \phi^2 (l) -r^2  ) \\
&  &
\hspace{3.5cm}  \times \left( l v_{ f }(x,r;l) -l^{-1}v_{f_0 } (x,r)+ l^{-4/3}\tau (l) \bigtriangleup  v_{f_0  }(x,r)\right)  
\end{eqnarray*}
implies (\ref{2.4}). Theorem is proven. \hfill $\Box$
\medskip

The last theorem does not exhaust the possible singularities of the source terms. 
The next theorem  gives behavior of the solution as $t\rightarrow   0$ if the source term is more singular.

\begin{theorem}
\label{T2.2new}
Assume that 
$ f(x,t) \in C^{[\frac{n}{2}]+4}({\mathbb  R}^n\times (0,\infty)) $ and that with   number $a \in (2,8/3)$  one has $  t^{  a }f(x,t)\in C({\mathbb  R}^n\times [0,\infty))$ 
and
\[
|\partial_x^\alpha f(x,t)| + |t \partial_t \partial_x^\beta  f (x,t)| \leq C_\alpha t^{ -a }  \qquad \mbox{for all}\quad   t\in (0,T) ,\,\, x \in {\mathbb R}^n, 
\]
and for every $\alpha  $, $ \beta $, $|\alpha | \leq  [\frac{n}{2}]+4$, $|\beta  | \leq  [\frac{n}{2}]+1$.  Denote $\displaystyle f_0(x) : =  \lim_{t\rightarrow 0} t^{a}f(x,t)$  and suppose that 
 with some $\varepsilon >0$ for the functions $f=f(x,t) $   and $f_0 =f_0(x) \in C^{ [\frac{n}{2}]+4}({\mathbb  R}^n) $
the following inequality is fulfilled:
\[
|\partial_x^\alpha \left( t  f(x,t) -t^{1-a}f_0(x) \right) | + |t \partial_t \partial_x^\beta  \left( t  f(x,t) -t^{1-a}f_0(x) \right)| \leq C_\alpha t^{\varepsilon  -1 }  
\,\, \mbox{for all}\,\,   t\in (0,T) ,\,\, x \in {\mathbb R}^n, 
\]
and for every $\alpha  $, $ \beta $, $|\alpha | \leq  [\frac{n}{2}]+2$, $|\beta  | \leq  [\frac{n}{2}]+1$. Denote $\tau (t):=  (3-a)^{-1}(2-a)^{-1}  t^{3-a}  $.

Then the solution $\psi =\psi (x,t) $ of the problem  
\begin{eqnarray}
\label{2.15}
\begin{cases}{
\vspace{0.2cm}  \psi_{tt} - t^{-4/3}\bigtriangleup    \psi +   2   t^{-1}          \psi_t = f (x,t),  \qquad t>0 ,\,\, x \in {\mathbb R}^n,\cr 
 \vspace{0.2cm}\displaystyle  \lim_{t \to 0} \left( t \psi   (x,t)- \frac{1}{(3-a)(2-a)} t^{3-a}f_0(x)\right)   =  0 \,, \quad x \in {\mathbb  R}^n ,\cr 
  \displaystyle \lim_{t \to 0} \left(  t \psi _t (x,t)+ \psi   (x,t)  - \frac{1}{ 2-a } t^{2-a}f_0(x)\right)
  =  0 \,, \quad x \in {\mathbb  R}^n , }
\end{cases}
\end{eqnarray} 
is given by 
\begin{eqnarray}
\label{2.16}
\psi  (x,t) 
& = & 
\frac{1}{t}f_0(x)\tau (t )+ \frac{1}{18t}   \int_{ 0}^{t} dl
\int_{ 0 }^{ \phi (t) - \phi (l) }  dr \, 
 (\phi^2 (t)  + \phi^2 (l) -r^2  ) \\
&  &
\hspace{3.5cm}  \times \left( l v_{ f }(x,r;l) -l^{1-a}v_{f_0 } (x,r)+ l^{-4/3}\tau (l)  \bigtriangleup  v_{ f_0  }(x,r)\right)  \nonumber \,.
\end{eqnarray}
\end{theorem}
\medskip

\noindent
{\bf Proof.} 
Consider the new unknown function $ w(x,t):=u-f_0(x)\tau (t ) $. 
Then
\begin{eqnarray*}
{\mathcal S}w(x,t)
& = &
tf(x,t) - {\mathcal S}(f_0(x)\tau (t) ) \\
& = &
tf(x,t)    -\left(t^{1-a}f_0(x)-t^{-4/3}\tau (t)  \bigtriangleup f_0(x) \right)\\
& = &
h(x,t),
\end{eqnarray*}
where we have denoted   
\[
h(x,t) := t\left( f(x,t) -t^{-a}f_0(x)\right) + t^{-4/3}\tau (t) \bigtriangleup f_0(x)\,.
\]
According to the condition of the theorem with some $\varepsilon >0 $ we have
\[
|\partial_x^\alpha h(x,t)| + |t \partial_x^\beta  h_t(x,t)| \leq C_\alpha t^{\varepsilon  -1 }  \qquad \mbox{for all}\quad   t\in (0,T) ,\,\, x \in {\mathbb R}^n, 
\]
$\alpha  $, $ \beta $, $|\alpha | \leq  [\frac{n}{2}]+2$, $|\beta  | \leq  [\frac{n}{2}]+1$, that allows us  to write representation (\ref{2.13}).
On the other hand, according to Theorem~\ref{T1.1}, the function $w=w(x,t) $ satisfies initial conditions (\ref{2.14}).
Consequently,
\begin{eqnarray*}
\lim_{t \to 0} (u  (x,t)- f_0(x)\tau (t ))
& = &   
\lim_{t \to 0}  w  (x,t)   =0 , \\
 \lim_{t \to 0} \left( u_t (x,t)-  f_0(x)\tau '(t)  \right)
 & =  &
\lim_{t \to 0}  w_t  (x,t) =  0 \,.
\end{eqnarray*}
For the function $\psi =\psi (x,t) = t^{-1}u (x,t) $ this implies the initial conditions 
\begin{eqnarray*}
\lim_{t \to 0} ( t \psi   (x,t)- f_0(x)\tau (t ))
& = &   
0 , \qquad 
 \lim_{t \to 0} \left(  t \psi _t (x,t)+ \psi   (x,t)  - f_0(x)\tau '(t)  \right)
  =  0 \,,
\end{eqnarray*}
which coincide with ones of (\ref{2.15}). To prove representation formula (\ref{2.16}), we note that
\begin{eqnarray*}
v_h (x,r;b) 
& = &
v_{t\left( f(x,t) -t^{-a}f_0(x)\right) + t^{-4/3}\tau (t) \bigtriangleup f_0(x)}(x,r;b)\\
& = &
v_{t   f(x,t) }(x,r;b) -v_{t^{1-a}f_0(x)} (x,r;b)+ v_{t^{-4/3}\tau (t) \bigtriangleup f_0(x) }(x,r;b)\\
& = &
b v_{ f }(x,r;b) -b^{1-a}v_{f_0 } (x,r)+ b^{-4/3}\tau (b) \bigtriangleup v_{f_0  }(x,r)\,.
\end{eqnarray*}
Then
\begin{eqnarray*}
w(x,t) 
& = &  
 \frac{1}{18}   \int_{ 0}^{t} dl
\int_{ 0 }^{ \phi (t) - \phi (l) }  dr \,  v_{h} (x,r ; l)  
 (\phi^2 (t)  + \phi^2 (l) -r^2  )\\
& = &  
 \frac{1}{18}   \int_{ 0}^{t} dl
\int_{ 0 }^{ \phi (t) - \phi (l) }  dr \, 
 (\phi^2 (t)  + \phi^2 (l) -r^2  ) \\
&  &
\qquad \times \left( l v_{ f }(x,r;l) -l^{1-a}v_{f_0 } (x,r)+ l^{-4/3}\tau (l) \bigtriangleup v_{f_0  }(x,r)\right)    \,.
\end{eqnarray*}
Thus, the following representation 
\begin{eqnarray*}
u (x,t) 
& = & 
f_0(x)\tau (t )+ \frac{1}{18}   \int_{ 0}^{t} dl
\int_{ 0 }^{ \phi (t) - \phi (l) }  dr \, 
 (\phi^2 (t)  + \phi^2 (l) -r^2  ) \\
&  &
\hspace{3.5cm}  \times \left( l v_{ f }(x,r;l) -l^{1-a}v_{f_0 } (x,r)+ l^{-4/3}\tau (l) \bigtriangleup v_{f_0  }(x,r)\right) 
\end{eqnarray*}
for the function $u=u(x,t)$ implies (\ref{2.16}). Theorem is proven. \hfill $\Box$
 
\bigskip

\noindent
{\bf The case of {\bf ($\bf \varphi_0$)}.} In this case $f=0$ and $  \varphi_1=0$. One can find in the literature different approaches for the construction of the solutions of the
 Fuchsian and non-Fuchsian partial differential equations. (See, e.g.  \cite{Mandai}, \cite{Parenti-Tahara}.) 
The next two lemmas give for $f=0$  behavior of the solutions of the equation  of (\ref{ivp}) near the point of singularity $t=0$ of the coefficients. 

\begin{lemma}
\label{L2.1}
For   $\varphi _0 \in C_0^{[\frac{n}{2}]+3} ({\mathbb R}^n) $    the function
\begin{eqnarray}
\label{4}
u (x,t)
& = & 
  v_{\varphi_0}  (x, 3t^{1/3})  - 3t^{1/3}   ( \partial_r   v_{\varphi_0} )  (x, 3t^{1/3})  
\end{eqnarray}
solves the problem
\[
 \cases{ \displaystyle  {\mathcal S}  u  = 0,  \qquad x \in {\mathbb R}^n,\quad t>0,\cr
 \displaystyle  \lim_{t\rightarrow 0}u (x,t) = \varphi_0 (x),  \qquad
 \displaystyle   \lim_{t\rightarrow 0}
\Big (
  u_t  (x,t) +3 t^{   - {1}/{3}  } \bigtriangleup \varphi_0 
   (x   )
  \Big )
=0,  \qquad x \in {\mathbb R}^n . }
\]
Here $v_{\varphi  }   ( x,3t^{1/3} ) $ is the value of the solution $v   ( x,r  ) $ to the Cauchy problem for the  wave equation,
$
v_{rr} -    \bigtriangleup v =0 $,  \,\,
$v(x,0) = \varphi  (x)$, \, $ v_{t } (x,0) =  0,
$
taken at the point $ ( x,r  )=( x,3t^{1/3})$. \end{lemma}
\medskip

\noindent
{\bf Proof.} We verify it by straightforward calculations.  
It is evident that 
\begin{eqnarray}
\label{deltau2}
\bigtriangleup u(x,t)
& = &
   \bigtriangleup v_{\varphi_0 }  (x, 3t^{1/3})- 3t^{1/3}  \left(   \partial_r  \bigtriangleup v_{\varphi_0 }   (x, r) \right)_{r=3t^{1/3}}.
\end{eqnarray}
Denote
 \begin{eqnarray*}
v_0(x,t)
  =  
 v_{\varphi_0 }  (x, 3t^{1/3}) ,      \qquad
v_1(x,t)
  =  
- 3t^{1/3}  \left(   \partial_r   v_{\varphi_0 }   (x, r) \right)_{r=3t^{1/3}} .  
\end{eqnarray*}
Then, for the derivatives $ \partial_t v_0(x,t)$ and $\partial_t^2 v_0(x,t) $ we have
\begin{eqnarray*}
\partial_t v_0(x,t)
& = &
t^{-2/3}\left(    \partial_r  v_{\varphi_0 }   (x, r) \right)_{r=3t^{1/3}},\\
\partial_t^2 v_0(x,t)
& = &
-\frac{2}{3}   t^{-5/3} \left(   \partial_r  v_{\varphi_0 }   (x, r) \right)_{r=3t^{1/3}}
+t^{-4/3}  \left(    \partial_r ^2  v_{\varphi_0 }   (x, r) \right)_{r=3t^{1/3}}.
\end{eqnarray*}
At the mean time for the derivatives $ \partial_t v_1(x,t)$ and $\partial_t^2 v_1(x,t) $ we have
\begin{eqnarray*}
\partial_t v_1(x,t)
& = &
-t^{-{2}/{3}}\left(  \partial_r v_{\varphi_0 }   (x, r) \right)_{r=3t^{1/3}}
-3t^{-{1}/{3}}\left(  \partial_r^2  v_{\varphi_0 }   (x, r) \right)_{r=3t^{1/3}},\\
\partial_t^2 v_1(x,t)
& = &  
   \frac{2}{3}  t^{-{5}/{3}}
\left(  \partial_r  v_{\varphi_0 }   (x, r) \right)_{r=3t^{1/3}}   
-3t^{-1} \left(   \partial_r^3  v_{\varphi_0 }   (x, r) \right)_{r=3t^{1/3}} .
\end{eqnarray*}
Hence, for the first order derivative $\partial_t  u(x,t)  $ and for the second  order derivative $\partial_t^2 u(x,t)  $ we have
\begin{eqnarray}
\partial_t u(x,t)
& = &
-3t^{-{1}/{3}}\left(   \partial_r^2  v_{\varphi_0 }   (x, r) \right)_{r=3t^{1/3}}, \nonumber \\
\label{utt2}
\partial_t^2 u(x,t)
& = &
   t^{-{4}/{3}}  \left(   \partial_r^2  v_{\varphi_0 }   (x, r) \right)_{r=3t^{1/3}}
 -3t^{-1}     \left(  \partial_r^3  v_{\varphi_0 }   (x, r) \right)_{r=3t^{1/3}},
\end{eqnarray}
respectively. Consequently, using (\ref{deltau2}), (\ref{utt2}), and the definition of $v_\varphi $ we obtain
\begin{eqnarray*}
\partial_t^2 u(x,t) - t^{-4/3}\bigtriangleup u(x,t) 
&  =  &
 t^{- {4}/{3}}  \left(   \partial_r^2  v_{\varphi_0 }   (x, r) \right)_{r=3t^{1/3}}
 -3t^{-1}     \left(   \partial_r^3  v_{\varphi_0 }   (x, r) \right)_{r=3t^{1/3}}\\
 &  &
 - t^{- {4}/{3}} \left( \bigtriangleup v_{\varphi_0 }  (x, r)\right)_{r=3t^{1/3}} + 3t^{-1}  \left(   \partial_r \bigtriangleup v_{\varphi_0 } (x, r) \right)_{r=3t^{1/3}}\\
&  =  &
 t^{- {4}/{3}}  \left(  \partial_r^2  v_{\varphi_0 }   (x, r) - \bigtriangleup v_{\varphi_0 }  (x, r) \right)_{r=3t^{1/3}}
 \\
 &  &
-3t^{-1}     \left(  \partial_r  \left(  \partial_r^2  v_{\varphi_0 }   (x, r)   
-   \bigtriangleup v_{\varphi_0 }  (x, r) \right)\right)_{r=3t^{1/3}}\\
 & = &
 0\,.
\end{eqnarray*}
Thus,     the function $u=u(x,t)$ solves the equation   $ u_{tt} (x,t) - t^{-4/3}\bigtriangleup  u(x,t) =0$. Lemma is proven. \hfill $\Box$

\begin{corollary}
\label{C2.4}
The function $\psi =t^{-1} u(x,t)$ solves the problem (\ref{ivp}) with $\varphi _1=0$ and with  $f=0$, that is 
\[
\cases{   \psi_{tt} (x,t) - t^{-4/3}\bigtriangleup  \psi (x,t) +2t^{-1}   \psi_t (x,t) =0\,, \cr
\displaystyle  \lim_{t\rightarrow 0}
  t \psi    (x,t)  =\varphi _0 , \qquad  \lim_{t\rightarrow 0}
\Big (
  t \psi _t  (x,t) +\psi    (x,t) +3 t^{   -\frac{1}{3}  } \bigtriangleup \varphi_0 
   (x   )
  \Big )
=0 ,  \qquad x \in {\mathbb R}^n. }
\]
\end{corollary}
In particular, the corollary  shows that for the given dimension $n \in {\mathbb N}$ Huygens' principle is  valid for some 
particular waves
propagating in the Einstein~\&~de~Sitter  model of the universe if and only if  it is valid for the waves
propagating in Minkowski spacetime (cf. with \cite{Sonego-Faraoni},  \cite{YagTricomi},   \cite{Yagdjian-Galstian}).

\bigskip

\noindent
{\bf The case of {\bf ($\bf \varphi_1$)}.} In this case $f=0$ and $  \varphi_0=0$.

\begin{lemma}  
\label{L2.3}
For   $\varphi _1 \in C_0^{[\frac{n}{2}]+2} ({\mathbb R}^n) $    the function
\begin{eqnarray}
\label{wtv}
u(x,t)
& = &
\, t\frac{3}{2}
\int_{0}^{1}  v_{\varphi _1}   ( x,\phi (t)  s  )  (1-s^2) ds, \quad x \in {\mathbb R}^n, \,\, t>0\,, 
\end{eqnarray}
solves the problem
\[
 \cases{ \displaystyle  {\mathcal S}  u  = 0,  \qquad x \in {\mathbb R}^n, \,\, t>0\,, \cr
 \displaystyle  \lim_{t\rightarrow 0}u (x,t) = 0,  \qquad
 \displaystyle   \lim_{t\rightarrow 0} 
    u_t  (x,t)   
=  \varphi_1 (x),  \qquad x \in {\mathbb R}^n.}
\]
Here $v_{\varphi  }   ( x,\phi (t)  s  ) $ is the value of the solution $v   ( x,r  ) $ to the Cauchy problem for the  wave equation,
$
v_{rr} -    \bigtriangleup v =0 $,  \,\,
$v(x,0) = \varphi  (x)$, \, $ v_{t } (x,0) =  0,
$
taken at the point $ ( x,r  )=( x,\phi (t)  s  )$, while $\phi (t)= 3t^{1/3}$. 
\end{lemma}
\medskip

\noindent
{\bf Proof.} We prove the lemma by straightforward calculations. We have
\begin{eqnarray*}
u(x,t)
& = &
\, t\frac{3}{2}
\int_{0}^{1}  v_{\varphi _1}   ( x,\phi (t)  s  )  (1-s^2) ds 
  =  
\frac{1}{18}
\int_{0}^{\phi (t)}  v_{\varphi _1}   ( x,  r  )  (\phi^2 (t)-r^2) dr. 
\end{eqnarray*}
 For the first order derivative we derive
\begin{eqnarray*}
 \partial_t  u(x,t)
& = &
 \partial_t \frac{1}{18}
\int_{0}^{\phi (t)}  v_{\varphi _1}   ( x,  r  )  (\phi^2 (t)-r^2) dr 
  =  
 \frac{1}{3}t^{- {1}/{3}} 
\int_{0}^{\phi (t)}  v_{\varphi _1}   ( x,  r  )   dr\,, 
\end{eqnarray*}
while for the second order derivative using the last equation and integration by parts we obtain
\begin{eqnarray*}
 \partial_t^2 u(x,t)
& = &
- \frac{1}{9}t^{- {4}/{3}} 
\int_{0}^{\phi (t)}  v_{\varphi _1}   ( x,  r  )   dr +\frac{1}{3}t^{-1} 
v_{\varphi _1}   ( x,  \phi (t)  ) \\
& = &
-  \frac{1}{9}t^{- {4}/{3}}  \left( v_{\varphi _1}   ( x,  r  )    r\Bigg|_{0}^{\phi (t)}-
\int_{0}^{\phi (t)} r \left(  \partial_r v_{\varphi _1}   \right)  ( x,  r  )   dr \right)
 +\frac{1}{3}t^{-1} 
v_{\varphi _1}   ( x,  \phi (t)  ) \\
& = &
-  \frac{1}{9}t^{- {4}/{3}}  \left( v_{\varphi _1}   ( x,  \phi (t)  )  \phi (t) -
\int_{0}^{\phi (t)} r \left(  \partial_r v_{\varphi _1}   \right)   ( x,  r  )   dr \right)
 +\frac{1}{3}t^{-1} 
v_{\varphi _1}   ( x,  \phi (t)  ) \,.
\end{eqnarray*}
Consequently,
\begin{eqnarray}
\label{2.21}
 \partial_t^2 u(x,t)
& = & 
\frac{1}{9}t^{- {4}/{3}} \int_{0}^{\phi (t)} r \left(  \partial_r v_{\varphi _1}   \right)   ( x,  r  )   dr \,.
\end{eqnarray}
At the same time, we have
\begin{eqnarray}
\label{2.22}
\bigtriangleup u(x,t)
& = &
\frac{1}{18}
\int_{0}^{\phi (t)} \bigtriangleup  v_{\varphi _1}    ( x,  r  )  (\phi^2 (t)-r^2) dr\,. 
\end{eqnarray}
Then equations (\ref{2.21}) and (\ref{2.22}) imply 
\begin{eqnarray*}
&  &
 u_{tt}(x,t) - t^{- {4}/{3}}\bigtriangleup u(x,t)\\
& = &
 \frac{1}{9}t^{- {4}/{3}}  \left( \frac{r^2}{2}\left(  \partial_r v_{\varphi _1}   \right) ( x,  r  )\Bigg|_{0}^{\phi (t)}-
\int_{0}^{\phi (t)} \frac{r^2}{2}\left(  \partial_r^2 v_{\varphi _1}   \right)   ( x,  r  )   dr \right)\\
&  &
- t^{- {4}/{3}}\frac{1}{18}
\int_{0}^{\phi (t)}  \bigtriangleup  v_{\varphi _1}  ( x,  r  )  (\phi^2 (t)-r^2) dr\\
& = &
 \frac{1}{9}t^{- {4}/{3}}  \left( \frac{\phi^2 (t)}{2} \left(  \partial_r v_{\varphi _1}   \right)  ( x, \phi (t)  ) -
\int_{0}^{\phi (t)} \frac{r^2}{2}\left(  \partial_r^2 v_{\varphi _1}   \right)  ( x,  r  )   dr \right)\\
&  &
- t^{- {4}/{3}}\frac{1}{18}
\int_{0}^{\phi (t)} \bigtriangleup  v_{\varphi _1}  ( x,  r  )  (\phi^2 (t)-r^2) dr\\
& = &
 \frac{1}{18}t^{- {4}/{3}}    \phi^2 (t)  \left(  \partial_r v_{\varphi _1}   \right)  ( x, \phi (t)  ) -
\frac{1}{18}t^{- {4}/{3}} \int_{0}^{\phi (t)}  r^2 \left(  \partial_r^2 v_{\varphi _1}   \right) ( x,  r  )   dr  \\
&  &
- t^{- {4}/{3}}\frac{1}{18}
\int_{0}^{\phi (t)}  \bigtriangleup  v_{\varphi _1} ( x,  r  )  (\phi^2 (t)-r^2) dr\\
& = &
 \frac{1}{2}t^{- {2}/{3}}   \left(  \partial_r v_{\varphi _1}   \right)   ( x, \phi (t)  )    
-  \frac{1}{2}t^{- {2}/{3}} 
\int_{0}^{\phi (t)}   \bigtriangleup  v_{\varphi _1}     ( x,  r  )   dr\,.
\end{eqnarray*}
The definition of the function  $v_{\varphi _1}$ suggests that the function $u=u(x,t)$ solves the equation:
\begin{eqnarray*}
 u_{tt}(x,t) - t^{- {4}/{3}}\bigtriangleup u(x,t)
& = &
 \frac{1}{2}t^{- {2}/{3}}  \left\{  \left(  \partial_r v_{\varphi _1}   \right)  ( x, \phi (t)  )    
-   
\int_{0}^{\phi (t)} \left(  \partial_r^2 v_{\varphi _1}   \right)   ( x,  r  )   dr \right\}\\
 & = &
 0\,.
\end{eqnarray*}
Finally we verify the second initial condition by means of the l'Hospital's rule:
\begin{eqnarray*}
\lim_{t \to 0}  u_t(x,t)
 = 
\lim_{t \to 0}  \frac{1}{3}t^{- {1}/{3}} 
\int_{0}^{\phi (t)}  v_{\varphi _1}   ( x,  r  )   dr = \lim_{t \to 0}  
v_{\varphi _1}   ( x,  \phi (t)  )    = v_{\varphi _1}   ( x,  0 ) =\varphi _1 ( x) \,.
\end{eqnarray*}
Lemma is proven. \hfill $\Box$
\begin{corollary}
\label{C2.6}
The function $\psi =t^{-1} u(x,t)$ solves the problem (\ref{ivp}) with $\varphi _0=0$ and without source term $f=0$, that is 
\begin{eqnarray*}
\begin{cases}{
 \psi_{tt} - t^{-4/3}\bigtriangleup    \psi +   2   t^{-1}          \psi_t = 0,  \qquad t>0 ,\,\, x \in {\mathbb R}^n,\cr 
 \displaystyle   \lim_{t\rightarrow 0}\, t \psi  (x,t) = 0, \quad   
\displaystyle   
\lim_{t\rightarrow 0} 
  \left(  t \psi_t  (x,t) + \psi  (x,t)   \right)
=  \varphi_1 (x),   \,\, x \in {\mathbb  R}^n .}
\end{cases}
\end{eqnarray*} 
\end{corollary}
The last corollary completes the proof of Theorem~\ref{T1.1}. \hfill $\square$

In particular, Corollary~\ref{C2.4} and Corollary~\ref{C2.6} show that, because of the integration in the formula (\ref{wtv}), for all $n \in {\mathbb N}$ Huygens' principle is not valid for waves
propagating in the Einstein~\&~de~Sitter  model of the universe, unless $\varphi _1=0$ and $f=0$.
\bigskip

\section{$\bf L^p-L^q$ estimates}
\label{S3}
\setcounter{equation}{0}
\renewcommand{\theequation}{\thesection.\arabic{equation}}

The representation formula (\ref{1.15}) of Theorem~\ref{T1.1} can be used to reproduce for the solutions of the wave equation in Einstein~\&~de~Sitter spacetime
 some important properties
which possess the  solutions of the wave equation in Minkowski spacetime. Among them there are estimates of the norm of solution in various functional spaces, such as $L^p$, Sobolev spaces,
Besov spaces and others. These estimates provide a useful tool to prove local and global in time existence theorems \cite{Shatah},  \cite{Strauss}, \cite{YagTricomi_GE},
\cite{YagTricomi_JMAA}. 

In this short note we derive such estimates in the Lebesgue spaces only. First we remind these estimates.
If $n \geq 2$, then for the solution $v = v (x,t)$ of the Cauchy problem for the wave equation in Minkowski spacetime
\begin{eqnarray}
\label{eq_0}
v_{tt}-  \bigtriangleup v =0, \quad v(x,0)= \varphi (x)  , \quad v_t(x,0)=0\,,
\end{eqnarray}
with $\varphi (x) \in C_0^\infty({\mathbb R}^n)$ one has (see, e.g., \cite{Brenner}, \cite{Pecher}) the following so-called $ L^p-L^q$ decay estimate
\begin{eqnarray}
\label{3.1}
&  & 
\| 
(-\bigtriangleup )^{-s} v (\cdot ,t) \|_{ { L}^{q} ({\mathbb R}^n)  }
\le C 
t^{2s-n(\frac{1}{p}-\frac{1}{q})}\|\varphi  \|_{ { L}^{p}({\mathbb R}^n)  }  \quad \mbox{\rm for all} \quad  t >0,
\end{eqnarray}
provided that \, $s\ge 0$, $1<p\le 2$, $\frac{1}{p}+\frac{1}{q}=1$, \,
and\,
$\frac{1}{2}(n+1)(\frac{1}{p}-\frac{1}{q})  \le 2s 
\le n (\frac{1}{p}-\frac{1}{q})$. 

Then, for the solution $v = v (x,t)$ of the Cauchy problem for the wave equation 
\begin{eqnarray}
\label{eq_1}
v_{tt}-  \bigtriangleup v =0, \quad v(x,0)= 0  , \quad v_t(x,0)=\varphi (x)\,,
\end{eqnarray}
there is the $ L^p-L^q$ estimate  
\begin{eqnarray}
\label{3.2}
&  &
\| 
(-\bigtriangleup )^{-s} v (\cdot ,t) \|_{ { L}^{q} ({\mathbb R}^n)  }
\le C\; 
t^{2s+1-n(\frac{1}{p}-\frac{1}{q})}\|\varphi  \|_{ { L}^{p}({\mathbb R}^n)  }    \quad \mbox{\rm for all} \quad  t >0,  
\end{eqnarray}
under the conditions $s\ge 0$, $1<p\le 2$, $\frac{1}{p}+\frac{1}{q}=1$, 
  and
$\frac{1}{2}(n+1)(\frac{1}{p}-\frac{1}{q}) - 1 \le 2s 
\le n (\frac{1}{p}-\frac{1}{q})$. 
\medskip

\noindent
{\bf The case of {\bf ($\bf \varphi _0$)}.}
According to Theorem~\ref{T1.1}, for the problem with $\varphi_1 =0$ and $f=0 $ the  function   $ \psi =\psi (x,t)$ can be represented as follows:  
\begin{eqnarray}
\label{3.5}
\psi (x,t) 
 & = & t^{  -1  } v_{\varphi_0}  (x, 3t^{1/3})- 3t^{-2/3}     \left( \partial_t v_{\varphi_0} \right)    (x, 3t^{1/3}) \,.
\end{eqnarray}
Here for $\varphi_0 \in C_0^\infty ({\mathbb R}^n)$ 
the function $v_{\varphi_0}  (x, 3t^{1/3})$  coincides with the value $v(x, 3t^{1/3}) $ 
of the solution $v(x,t)$ of the Cauchy problem (\ref{eq_0}). 
Hence for \, $s\ge 0$\, by means of application of  (\ref{3.1}) we obtain 
\begin{eqnarray*}
&  & 
\| 
(-\bigtriangleup )^{-s} v_{\varphi_0}  (\cdot , 3t^{1/3})\|_{ { L}^{q} ({\mathbb R}^n)  }
\le C 
t^{\frac{1}{3}\left( 2s-n(\frac{1}{p}-\frac{1}{q}) \right)}\|\varphi_0   \|_{ { L}^{p}({\mathbb R}^n)  } , \,\, t >0. 
\end{eqnarray*}
To estimate the second term of (\ref{3.5}) we apply (\ref{3.2})  with \, $s\ge 0$\, :
\begin{eqnarray*}
&  & 
\| 
(-\bigtriangleup )^{-s}  (\partial_r  v_{\varphi_0} )     (\cdot , 3t^{1/3})\|_{ { L}^{q} ({\mathbb R}^n)  }
\le C 
t^{\frac{1}{3}\left( 2s+1-n(\frac{1}{p}-\frac{1}{q}) \right)}\| \bigtriangleup \varphi_0   \|_{ { L}^{p}({\mathbb R}^n)  } , \,\, t >0\,,
\end{eqnarray*}
provided that $\frac{1}{2}(n+1)(\frac{1}{p}-\frac{1}{q}) - 1 \le 2s 
\le  n (\frac{1}{p}-\frac{1}{q})$. Consequently, if  $s\geq 0$,  $1<p\le 2$, $\frac{1}{p}+\frac{1}{q}=1$,   and
$\frac{1}{2}(n+1)(\frac{1}{p}-\frac{1}{q}) \le 2s 
\le n (\frac{1}{p}-\frac{1}{q})$, then for the problem with $\varphi_1 =0$ and $f=0 $ we obtain
\begin{eqnarray*}
\| 
(-\bigtriangleup )^{-s} \psi (\cdot ,t)\|_{ { L}^{q} ({\mathbb R}^n)  }
& \le  & 
C 
t^{-1+\frac{1}{3}\left( 2s-n(\frac{1}{p}-\frac{1}{q}) \right)}\|\varphi_0  \|_{ { L}^{p}({\mathbb R}^n)  } + C 
t^{-\frac{2}{3}}t^{\frac{1}{3}\left( 2s+1-n(\frac{1}{p}-\frac{1}{q}) \right)}\|\bigtriangleup \varphi_0  \|_{ { L}^{p}({\mathbb R}^n)  } \nonumber \\ 
& \le  & C 
t^{\frac{1}{3}\left( 2s-n(\frac{1}{p}-\frac{1}{q}) \right)}\left( t^{-1}\|\varphi_0  \|_{ { L}^{p}({\mathbb R}^n)  } + 
t^{-\frac{1}{3}} \|\bigtriangleup \varphi_0  \|_{ { L}^{p}({\mathbb R}^n)  } \right), \quad t >0. 
\end{eqnarray*}
Thus, we have proven the following proposition.
\begin{proposition}
\label{P3.1}
Suppose that  $s\geq 0$,  $1<p\le 2$, $\frac{1}{p}+\frac{1}{q}=1$,   and
$\frac{1}{2}(n+1)(\frac{1}{p}-\frac{1}{q})  \le 2s 
\le n (\frac{1}{p}-\frac{1}{q})$. Then the solution $ \psi = \psi (x,t)$ to the problem  
\begin{eqnarray*}
\begin{cases}{
 \psi_{tt} - t^{-4/3}\bigtriangleup    \psi +   2   t^{-1}          \psi_t = 0,  \qquad t>0 ,\,\, x \in {\mathbb R}^n,\cr 
 \displaystyle   \lim_{t\rightarrow 0}\, t \psi  (x,t) = \varphi_0 (x), \quad   
\displaystyle   
\lim_{t\rightarrow 0} 
  \left(  t \psi_t  (x,t) + \psi  (x,t)+3 t^{ - {1}/{3}} \bigtriangleup \varphi_0   (x   )  \right)
=  0,   \quad x \in {\mathbb  R}^n , }
\end{cases}
\end{eqnarray*} 
with $\varphi_0 \in   C_0^\infty({\mathbb R}^n) $ satisfies the following estimate
\begin{eqnarray}
\| 
(-\bigtriangleup )^{-s} \psi (\cdot ,t)\|_{ { L}^{q} ({\mathbb R}^n)  }
& \le  &C 
t^{\frac{1}{3}\left( 2s-1-n(\frac{1}{p}-\frac{1}{q}) \right)}\left( t^{-\frac{2}{3}}\|\varphi_0  \|_{ { L}^{p}({\mathbb R}^n)  } + 
 \|\bigtriangleup \varphi_0  \|_{ { L}^{p}({\mathbb R}^n)  } \right), \quad t >0, 
\end{eqnarray}
with the constant $C$ independent of $\varphi_0 $.
\end{proposition}

\noindent
{\bf The case of {\bf ($\bf \varphi _1$)}.}
For the problem with $\varphi_0 =0$ and $f=0 $  
the function  $\psi =\psi (x,t)$  due  to Theorem~\ref{T1.1} can be represented as follows:
\begin{eqnarray*} 
\psi  (x,t)
& = &
 \frac{3}{2} 
\int_{0}^1   v_{\varphi _1 } (x, \phi (t) s) 
(1-s^2) ds , \quad x \in {\mathbb R}^n, \,\, t>0\,.
\end{eqnarray*}
Then we obtain for $s $, $n$, $p$, and $q$ such that $  2s-n(\frac{1}{p}-\frac{1}{q}) >-1$,  the following estimate
\begin{eqnarray*} 
\| (-\bigtriangleup )^{-s} \psi (\cdot ,t)\|_{ { L}^{q} ({\mathbb R}^n)  }
& \leq  &
 \frac{3}{2} 
\int_{0}^1  \| 
(-\bigtriangleup )^{-s}  v_{\varphi _1 } (\cdot , \phi (t) s)\|_{ { L}^{q} ({\mathbb R}^n)  } 
(1-s^2) ds  \\
& \leq  &
 C 
\int_{0}^1  t^{\frac{1}{3}\left( 2s-n(\frac{1}{p}-\frac{1}{q}) \right)}s^{ \left( 2s-n(\frac{1}{p}-\frac{1}{q}) \right)}\|\varphi_1    \|_{ { L}^{p}({\mathbb R}^n)  } 
(1-s^2) ds \\
& \leq  &
C t^{\frac{1}{3}\left( 2s-n(\frac{1}{p}-\frac{1}{q}) \right)}\|\varphi _1   \|_{ { L}^{p}({\mathbb R}^n)  }
\int_{0}^1  s^{  2s-n(\frac{1}{p}-\frac{1}{q})} 
(1-s^2) ds \,.
\end{eqnarray*}
Thus, in this case we have proven the following proposition.
\begin{proposition}
\label{P3.2}
Suppose that  $s\geq 0$,  $1<p\le 2$, $\frac{1}{p}+\frac{1}{q}=1$, $  2s-n(\frac{1}{p}-\frac{1}{q}) >-1$,  and
$\frac{1}{2}(n+1)(\frac{1}{p}-\frac{1}{q}) \le 2s 
\le n (\frac{1}{p}-\frac{1}{q})$. Then the solution $ \psi = \psi (x,t)$ to the problem  
\begin{eqnarray*}
\begin{cases}{
 \psi_{tt} - t^{-4/3}\bigtriangleup    \psi +   2   t^{-1}          \psi_t = 0,  \qquad t>0 ,\,\, x \in {\mathbb R}^n,\cr 
 \displaystyle   \lim_{t\rightarrow 0}\, t \psi  (x,t) = 0, \quad   
\displaystyle   
\lim_{t\rightarrow 0} 
  \left(  t \psi_t  (x,t)+ \psi  (x,t)  \right)
=  \varphi_1 (x),   \quad x \in {\mathbb  R}^n , }
\end{cases}
\end{eqnarray*} 
with $\varphi_1 \in   C_0^\infty({\mathbb R}^n) $ satisfies the following estimate
\begin{eqnarray} 
\| (-\bigtriangleup )^{-s} \psi (\cdot ,t)\|_{ { L}^{q} ({\mathbb R}^n)  }
& \leq  &  
C t^{\frac{1}{3}\left( 2s-n(\frac{1}{p}-\frac{1}{q}) \right)}\|\varphi_1    \|_{ { L}^{p}({\mathbb R}^n)  }, \quad   t>0\,,
\end{eqnarray}
with the constant $C$ independent of $\varphi_1 $.
\end{proposition}
\medskip

\noindent
{\bf The case of {\bf ($\bf f$)}.}
According to Theorem~\ref{T1.1}, for the problem with $\varphi_0 =0$ and $\varphi_1 =0 $  
the function    $\psi =\psi (x,t)$  can be represented as follows:
\begin{eqnarray*}
\psi (x,t)
& = & 
\frac{3}{2} t^2 \int_{ 0}^{1} db
\int_{ 0 }^{ 1 - b^{1/3} }  d\tau \,  bv_f (x,3t^{1/3}  \tau  ; tb)  
 \big(1   + b^{2/3}-\tau ^2  \big) \,.
\end{eqnarray*}
Consequently, for the problem with $\varphi_0 =0$, $\varphi_1 =0 $, and the function $f$ satisfying conditions of the theorem, by application of (\ref{3.1}) we obtain
\begin{eqnarray*}
&  &
\| (-\bigtriangleup )^{-s} \psi (\cdot ,t)\|_{ { L}^{q} ({\mathbb R}^n)  } \\
& \leq  &
\frac{3}{2} t^2 \int_{ 0}^{1} db
\int_{ 0 }^{ 1 - b^{1/3} }  d\tau  \,  b \| (-\bigtriangleup )^{-s} v_f (\cdot ,3t^{1/3}  \tau  ; tb) \|_{ { L}^{q} ({\mathbb R}^n)  } 
 \big(1   + b^{2/3}-\tau ^2  \big) \\
& \leq  &
C t^2 \int_{ 0}^{1} db
\int_{ 0 }^{ 1 - b^{1/3} }  d\tau  \,  b  t^{\frac{1}{3}\left( 2s-n(\frac{1}{p}-\frac{1}{q}) \right)}
\tau ^{  2s-n(\frac{1}{p}-\frac{1}{q}) }\|f(\cdot ,tb)  \|_{ { L}^{p}({\mathbb R}^n)  }  
 \big(1   + b^{2/3}-\tau ^2  \big)\,.
\end{eqnarray*}
For $a= 2s-n(\frac{1}{p}-\frac{1}{q}) >-1 $ one has
\begin{eqnarray*}
 \int_{ 0 }^{ 1 - b^{1/3} } 
\tau ^{ a} 
 \big(1   + b^{2/3}-\tau ^2  \big) \,  d\tau=  \frac{2 }{(a+1)(a+3)}\left(1-b^{1/3}\right)^{a+1} \left(1+(a+1) b^{1/3}+b^{2/3}\right).  
\end{eqnarray*}
Hence,
\begin{eqnarray*}
&  &
\| (-\bigtriangleup )^{-s} \psi (\cdot ,t)\|_{ { L}^{q} ({\mathbb R}^n)  } \\
& \leq  &
C   t^{2+ \frac{1}{3}\left( 2s-n(\frac{1}{p}-\frac{1}{q}) \right)}\int_{ 0}^{1}   b \|f(\cdot ,tb)  \|_{ { L}^{p}({\mathbb R}^n)  } \,db
\int_{ 0 }^{ 1 - b^{1/3} } 
\tau ^{ 2s-n(\frac{1}{p}-\frac{1}{q})} 
 \big(1   + b^{2/3}-\tau ^2  \big) \,  d\tau  \\
& \leq  &
C_{n,p,q,s}   t^{2+ \frac{1}{3}\left( 2s-n(\frac{1}{p}-\frac{1}{q}) \right)}\int_{ 0}^{1}   b \|f(\cdot ,tb)  \|_{ { L}^{p}({\mathbb R}^n)  }  
  \left(1-b^{1/3}\right)^{a+1} \left(1+(a+1) b^{1/3}+b^{2/3}\right) \,db  \\
& \leq  &
C_{n,p,q,s}   t^{2+ \frac{1}{3}\left( 2s-n(\frac{1}{p}-\frac{1}{q}) \right)}\int_{ 0}^{1}   b \|f(\cdot ,tb)  \|_{ { L}^{p}({\mathbb R}^n)  }  
  \left(1-b^{1/3}\right)^{2s-n(\frac{1}{p}-\frac{1}{q})+1}  \,db.  
\end{eqnarray*}
Thus, in this case we have proven the following proposition.
\begin{proposition}
\label{P3.3}
Suppose that  $s\ge 0$, $1<p\le 2$, $\frac{1}{p}+\frac{1}{q}=1$,  $  2s-n(\frac{1}{p}-\frac{1}{q}) >-1$, and
$\frac{1}{2}(n+1)(\frac{1}{p}-\frac{1}{q})   \le 2s 
\le n (\frac{1}{p}-\frac{1}{q})$, and that the function $f$ satisfies conditions of Theorem~\ref{T1.1}. Then for the solution $ \psi = \psi (x,t)$ to the problem 
\begin{eqnarray*}
\begin{cases}{
 \psi_{tt} - t^{-4/3}\bigtriangleup    \psi +   2   t^{-1}          \psi_t = f (x,t),  \qquad t>0 ,\,\, x \in {\mathbb R}^n,\cr 
 \displaystyle   \lim_{t\rightarrow 0}\, t \psi  (x,t) = 0, \quad   
\displaystyle   
\lim_{t\rightarrow 0} 
  \left(  t \psi_t  (x,t) + \psi  (x,t) \right)
=  0,   \quad x \in {\mathbb  R}^n , }
\end{cases}
\end{eqnarray*}
the following estimate 
\begin{eqnarray*}
\| (-\bigtriangleup )^{-s} \psi (\cdot ,t)\|_{ { L}^{q} ({\mathbb R}^n)  } 
& \leq  &
C_{n,p,q,s}   t^{  \frac{1}{3}\left( 2s-n(\frac{1}{p}-\frac{1}{q}) \right)}\int_{ 0}^{t} \tau \|f(\cdot ,\tau )  \|_{ { L}^{p}({\mathbb R}^n)  }  
\,d\tau  
\end{eqnarray*} 
holds with the constant $C_{n,p,q,s}$ independent of $f$.
\end{proposition}

\bigskip

\noindent
{\bf ACKNOWLEDGMENTS} 
\medskip

This work was initiated during the first and third authors  visit Institute of  Mathematics of the University of Tsukuba in June 2008. The first and the third authors would like to express their gratitude to  the University of Tsukuba for  the financial support. They are  especially grateful to Prof.~Kajitani, Prof.~Wakabayashi,  and  Prof.~Isozaki for their hospitality.
Finally, the authors thank the referee for useful comments.

 \medskip 

\end{document}